# Internet of Threats Introspection in Dynamic Intelligent Virtual Sensing

**Victor R. Kebande**
Internet of Things and People
Research Center,
Malmö Universitet
Malmö, Sweden
victor.kebande@mau.se

**Joseph Bugeja**
Internet of Things and People
Research Center,
Malmö Universitet
Malmö, Sweden
joseph.bugeja@mau.se

**Jan A. Persson**
Internet of Things and People
Research Center,
Malmö Universitet
Malmö, Sweden
jan.a.persson@mau.se

## ABSTRACT

Continued ubiquity of communication infrastructure across Internet of Things (IoT) ecosystems has seen persistent advances of dynamic, intelligent, virtualised sensing and actuation. This has led to effective interaction across the connected ecosystem of "things". Furthermore, this has enabled the creation of smart environments that has created the need for the development of different IoT protocols that support the relaying of information across billions of electronic devices over the Internet. That notwithstanding, the phenomenon of virtual sensors that are supported by IoT technologies like Wireless Sensor Networks (WSNs), RFID, WIFI, Bluetooth, ZigBee, IEEE 802.15.4, etc., emulates physical sensors, and enables more efficient resource management through the dynamic allocation of virtual sensor resources. A distinctive example of this has been the proposition of the Dynamic Intelligent Virtual Sensors (DIVS). This DIVS concept is a novel proposition that allows sensing to be done by the use of logical instances through the use of labeled data. This allows for making accurate predictions during data fusion. However, a potential security attack on DIVS may end up providing false labels during the User Feedback Process (UFP), which may interfere with the accuracy of DIVS. This paper investigates the threat landscape in DIVS when employed in IoT ecosystems, in order to identify the extent to which the severity of these threats may hinder accurate prediction of DIVS in IoT, based on labeled data. The authors have conducted a threat introspection in DIVS from a security perspective.

## 1 INTRODUCTION

The emergence of the Internet of Things (IoT) and the need to disseminate effective services to human beings across IoT environments has paved the way for the physical world to be digitally connected. Sensing has been at the center of all these proliferations, however, the need to enforce the security of information for smart IoT environments, connected 'things' and systems like Industrial Control Systems (ICS), cyber-physical Systems (CPS) and the Supervisory Control and Data Acquisition (SCADA) networks has given rise to the considerations of IoT security. Most of the IoT devices currently do not have advanced security capabilities and given the continued increase of IoT device's capabilities, information produced by these devices has increased in volume and complexity over the years, effectively widening the threat landscape. Currently, IoT-based attacks seem to be channeled towards the control systems and the Critical Infrastructure Systems (CIS) that mainly comprise embedded IoT devices and systems. The main target, however, is the information that is produced and exchanged by these devices and the services rendered. It is imperative to note that there is a need to ensure the safety of this information. This is as this information may involve attributes that can be inadvertently used to compromise the overall resilience, security and privacy of an IoT system and its users.

Virtual sensors provide an abstraction of physical computing resources that are able to be adopted as logical representations across users which brings about effectiveness [1] on how sensor data is processed during data fusion. However, during machine learning process, there may exist security challenges that can interfere or change the fusing data. While virtual sensors provide cost-effective approaches that allow them to utilise nodes when only needed, the use of virtual sensing has triggered other alarming security challenges in IoT environments. The most infamous security challenge has been passive and active threats that exist in virtual sensors in IoT ecosystems [33]. Preliminary studies that have been conducted on virtual sensors [2-7] have mainly focused on how WSNs can be deployed in a virtualised environment in order to achieve sensing as a service (SaaS) but security of virtual sensors is least explored under these circumstances. This research has been motivated by the fact that the User-feedback Process (UFP) in DIVS could advertently be attacked through malicious inputs through labeled data in order to influence the behavior of virtual sensors. Still attackers could use virtual sensors to push malicious code into IoT devices [21]. Consequently, besides attacking IoT devices over prevailing threats, adversaries can use sensor instances to attack other interconnected sensors in the case of virtual sensors.

Therefore, the authors prioritize the security perspective of virtual sensors from the standpoint of identifying the threats in virtual sensors and suggesting research directions. Additionally, a discussion that has used the Dynamic Intelligent Virtual Sensor (DIVS) proposition as a basis has formed the focal part of this study. Consequently, the authors through this paper take a step to give an introspection on the risks posed by threats in virtual sensors in IoT environment. To bring out the problem explicitly, the authors consider a fictitious hypothetical scenario below that has mainly been used for illustration purposes.

*XYZ is a smart campus, that allows real-time activity detection in study rooms. A user's activity can be detected through the presence of a DIVS, which inputs sensor data like sound level, temperature, motion etc. X is a malicious user who has managed to interfere with the DIVS during the User Feedback Process (UFP). Also, X has been able to masquerade using false labels and has managed to mount multiple illegal sensor nodes with the same identities within the network and this has also enabled a total shut down of the smart cameras. X has been able to achieve this because it is possible in the UFP to achieve this for instance through pushing a button or input of data through a panel. Apart from that, information between other DIVS has been re-routed and dynamic services have been denied.*

Based on the aforementioned challenge in the hypothetical scenario, it is important to note that the existence of DIVS, acts as an open environment for IoT-based virtual sensor threats given that, at the time of writing, the security aspects of virtual sensors has not been explored.

**Contributions:** The authors give the contribution of this paper as follows:

1. Give an introspection of the threats in IoT in the perspective of DIVS dubbed Virtual Internet of Threats (VIoT);
2. Explore the possible IoT threats from an information security perspective using DIVS as a baseline;
3. Explore open security problems in virtual sensors, give a discussion on the propositions and suggest research direction worth taking.

**Organisation:** The remainder of this paper is structured as follows: Section 2 covers the background while Section 3 handles Virtual Internet of Threats (VIoT) adversarial model. After this, Section 4 explains the VIoT introspection in DIVS. This is followed by Section 5 that presents open security problems and future research directions. Next, Section 6 gives a discussion of the study. Finally, the paper concludes in Section 7 and make mention of future work.

## 2. BACKGROUND

This section provides background information on the following areas: virtual sensing in IoT, DIVS, and the need for DIVS security across IoT paradigms. Virtual sensing in IoT has been discussed in this paper to show how pertinent virtualisation is for the sensor networks in IoT. DIVS which is an intelligent virtual sensor forms the basis of the discussion in this paper. The need for DIVS security is discussed to show different technologies in WSN that face security challenges. These discussions are relevant to the study presented in the rest of the paper.

**Virtual sensing in IoT**
Research by [5] has highlighted that Virtualised Wireless Sensor Networks (VWSNs) are important for IoT paradigms if the paradigm is to achieve effective connectivity, scalability and cost saving approaches, which allows IoT users to get dedicated resources [8]. Apart from that, sensor virtualisation consists of instances running over applications on a sensor node that emulates a physical sensor. Additionally, virtual sensing in IoT is supported by several standards like ZigBee, Zwave, 6LowPAN, 802.11 and IEEE 802.15.4 [9-12]. Notably, research by [1] gives a different perspective of VWSN that is based on VWSN's implementation. These authors highlight that VWSNs can be implemented either at node-level or at network-level, where node-level allows multiple sensor tasks to be computed at a single sensor node concurrently. On the other hand, network-level virtualisation allows the formation of Virtual Sensor Network (VSN) by a subset of WSN nodes. This is apparent in the subsequent sections of this paper. Also, research by [13] has proposed a Dynamic Intelligent Virtual Sensor (DIVS) that can create abstraction layers over physical infrastructures to enable the logical instances to perform tasks, which has been discussed in the section to follow.

**Dynamic intelligent virtual sensor (DIVS)**
The Dynamic Intelligent Virtual Sensor (DIVS) which has been used as a preliminary study in this paper presents the notion of a virtual sensor that is deployed in a heterogeneous sensing environment. Based on Figure 1, DIVS has a machine learning component based on labelled instances. More precisely, DIVS uses heterogeneous sensor data which is able to undergo data fusion [13], [22]. Through the ability of online learning, DIVS is able to adjust with the changing nature of an IoT environment. Generally, DIVS creates an abstraction layer that overlays the physical infrastructure and the abstraction layer caters mainly of multiple logical instances [22]. Based on the availability of these logical instances, services can easily be managed or created based on available fusing data. An important aspect that forms part of the security consideration is that the user through the User Feedback Process (UFP) (see Fig 1) is able to provide input/feedback for learning purposes. The UFP in its entirety is not a secure communication process in DIVS. Figure 1 shows the DIVS data processing pipeline.

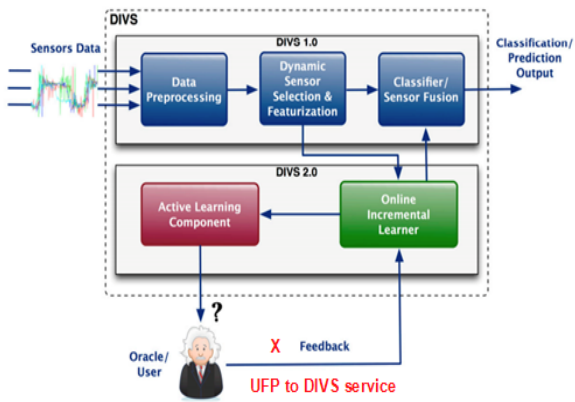

**Figure 1: DIVS Data processing pipeline [22]**

The UFP, marked X in the figure, is intended to be a process involving users to, on request by the DIVS or based on the user, provide information to improve the accuracy of the DIVS. The information is typically in the format of labeled data, i.e. the correct classification of the current state. Hence, the provision of false labels could rather quickly deteriorate the accuracy of prediction being made by the DIVS, i.e. the data fusion is modified through the online learning approach such that false predictions will be made, however, this channel faces a variety of threats. Of interest in this research is to explore the threat landscape from an information security perspective, using DIVS as a foundation. Also, it is important to explore how a DIVS attack can influence the accuracy of the DIVS.

**Need for DIVS security across IoT paradigms**
There is a need for enforcing the security technique of the DIVS in the IoT paradigm. This is because the common approach for the design of security solutions for sensors are generally related to the security functions that an IT product gives [34]. The authors of this paper emphasize the assumptions (threat model) that may be exploited by an attacker, owing to the fact, that the requirements of DIVS may change over time given the environment it is deployed in. In fact, the safety of virtual sensors should be supported by a number of architectural protocols, and the safety of this communication has also been backed up by the security of these protocols or technologies. That notwithstanding, the increased complexity of IoT threats and attacks has increased the need for sensor technology sensitization in order to ensure more secure communication. For example, 5G technology provides seamless connectivity due to low latency and high security through wireless communication, however, this technology requires to be authenticated [14], while ZigBee [18] uses low power wireless transmission and faces integrity and encryption issues. Radio Frequency Identification (RFID) which uses frequency waves requires encryption due to the susceptibility of integrity attacks [15]. Wireless Sensor Networks (WSNs) which use wireless technique to propagate requires encryption since information collected by sensors is sent to the server [16]. Wireless-Fidelity (Wi-Fi) that uses radio frequency signals requires authentication due to potential unauthorised access of information [19]. Message Queuing Telemetry Transport (MQTT) is a messaging protocol that uses a publish and subscribe model and it requires encryption techniques. MQTT has been used in the implementation of the DIVS concept, while IEEE 802.15.4 and 6LoWPAN for wireless requires authentication each respectively [20]. Finally, LoRaWAN which also uses long range wireless propagation mechanism requires encryption due to end devices being able to send messages to gateways [20].

In view of the foregoing, the DIVS concept [13] represents a virtual sensor that does not possess a security component and this study explores the extent to which DIVS may pose as a security threat or other threats that DIVS may face in an IoT environment. A successful attack at the DIVS concept could fulfil some of the adversarial motives that have been mentioned in the hypothetical scenario among others. Based on these shortcomings, the adversarial threat model is discussed next.

## 3   VIRTUAL INTERNET OF THREAT (VIOT) ADVERSARIAL THREAT MODEL

In this section, the authors highlight the adversarial threat model that is centered on the DIVS [13] based on the hypothetical scenario illustrated in Section 1. Insights on the DIVS concept are highlighted on a high-level standpoint (See Fig 1), which together with the threat model, that is presented in this section have been used to sum up the Internet of threat introspection discussion in this paper. The DIVS concept has been discussed in this section because it represents a virtual sensor that is susceptible to sensor threats from a security point of view.

**VIoT Attacker's Capability**
A threat is an act that can exploit security weaknesses in a system and exerts a negative impact on it. Sensor threats are active malicious actions that are more focused on compromising sensors through interference, leakage of information, draining sensor energy or through Denial of Service (DoS), etc [21]. Virtual sensors allow encapsulated layers of software to be able to provide services as a physical sensor, where sensor instances can perform tasks like physical sensors. However, the sensor instances are susceptible to threats just like in any virtualised environment – a virtualized environment involves virtual (rather than actual) computer hardware platforms, storage devices, and computer network resources [35]. Most of the sensor sources of threats result from the communication and interaction of the embedded physical and virtual processes of the devices. An assumption is made in our threat model that, the UFP in the DIVS is an insecure channel that involves a range of sensors, where some may be illegal sensor nodes or the information being relayed may travel over insecure channels (*See hypothetical scenario, Section 1*). Also, based on the hypothetical scenario, the authors assume that any user interacting with the DIVS service cannot be trusted, therefore raising trust issues. Additionally, the authors assume that integrity, confidentiality and authentication, which are some of the prime goals that are meant to be achieved in DIVS could be violated by a malicious user in the UFP. In this context, the threats could be targeted to the data fusion model through online learning based on the UFP.

**Threat Model**

The threat model, for the focus of analysis, is a culmination of the possibilities that may be experienced as a result of the execution of the DIVS (see Fig. 1) service to and from the oracle/user for labeled instances, which has been termed as the DIVS service-UFP. Given that virtual sensors are deployed in an uncontrolled, potentially open environment, the authors assume that the UFP has the potential of being captured or being tampered with by an adversary using a variety of techniques. While the existing literature [21], [32] has shown that sensors can resist being tampered with, e.g., through tamper-resistant packaging, our threat model focuses on the data transmitted between the oracle and the DIVS service. The authors argue that an adversary may be more interested to attack the UFP, resulting in inaccurate predictions, i.e, that can allow one to provide false labels in order to interfere with online learning of DIVS. This has also been based on the propositions of the Dolev-Yao intruder attacker model which is the basic foundation for adversary scenario [23]. The Dolev-Yao attacker model employs a set of rules that can outline the potential actions used by an attacker concerning information exchanged between parties during protocol execution. This foundation shows the duplex communication between two distinct nodes in a WSN during normal user-feedback process to the DIVS service. During this UFP to DIVS service, *A* as depicted in equation (i) could easily be transmitting an encrypted message {M} to *B* and *Z* could intercept {M} and re-encrypt to M ({M}) as is shown in (i), (ii), (iii), (iv) and (v) respectively.

$A_{(send)} \rightarrow_{(transmit\_the\_process)} B_{(receive)}: \{M\}_{(encrypted)}\_B$     (i)

$B_{(send)} \rightarrow_{(echo\_ACK)} A_{(receive)}: \{M\}\_A_{(encrypted)}$     (ii)

This could be intercepted, modified and rerouted easily, hence the need to create or have correlating aspects;

$Z_{(adversary)} \rightarrow_{(intercept)} B: \{M\}\_B_{(encrypted)}$     (iii)

$B_{(received\ from\ Z)} \rightarrow_{(echo\_ACK)} Z: \{M\}\_Z_{(encrypted)}$     (iv)

$M\ (Z \rightarrow A: \{M\}\_A)_{(Re\text{-}encrypted)}$     (v)

Considering the aforementioned, there is a possibility of an adversary interference with the UFP to limit the accurate prediction of the DIVS service. The assumption is that the possible attacks that may be directed to DIVS service hold some characteristics as follows:

- False labels to the DIVS could interfere with the online learning process by giving outputs other than the originally intended hence affecting the accuracy of DIVS prediction during data fusion.
- The UFP between on the DIVS service may be an insecure transmission mechanism through which an adversary is able to have full or partial control which may make him able to modify or tamper.
- An adversary can deny service to the DIVS communication channel which may interfere with the UFP.
- If a gatekeeper is tasked in managing the UFP, an adversary can attack and capture it which eventually may break the entire communication channel of the UFP.
- Still, an adversary over the UFP channel could obtain sensitive data in a malicious way that could violate data privacy.
- An adversary could use the sensor instances of the DIVS as instruments of launching sensor-based or other malicious attacks.

Based on the above-mentioned characteristics, there is a need to highlight the security goals that are aimed to be achieved by the DIVS architecture based on the UFP. The prioritisation of the security goals depends on the control environment and how the services are dispatched. These goals have been inclined towards the integrity of the information being transmitted in order to avoid the injection of false data, communication alteration, tampering between the users and the DIVS service, authenticity of transmitting parties, privacy and trust. These concerns mainly represent top concerns that can be shared across IoT-based systems.

## 4. VIRTUAL INTERNET OF THREAT (VIOT) INTROSPECTION IN DIVS

In this section, the authors highlight VIoT introspection approaches in DIVS as a contribution that has been given from a security perspective. This section has concentrated on showing how virtual sensors are susceptible to threats and attacks in DIVS. This is then followed by a discussion on virtual sensor threats, vulnerabilities and attacks. It is important to note that the VIoT discussion presented in this section is inclined to the initially described DIVS and based on the analogies of the hypothetical scenario (See Section 1).

**VIoT from a security perspective**

Virtual sensors that mainly emulate physical sensors represent the interaction with the target environment using specialised software. The software in this context is used to allow the sensing of various context-aware entities in order to have a virtualised representation that emulates the physical electronic sensor nodes, where mostly many activities are associated to the traditional WSNs [25], [26]. For example, the most effective way to manage a million sensors that are deployed in a smart IoT environment, or a smart city, to monitor people's activities in order to collect sensor data is to apply intelligent virtual sensors. This is a cost-effective exercise where an application can utilise virtual sensors or opportunistic sensing [24]. From an information security standpoint, the existence of configuration flaws or vulnerabilities in sensors allow an adversary to use virtual sensors as instruments of perpetrating attacks. This is because a number of resources end up being shared which in the long run opens the possibility of shred vulnerabilities.

The authors have explored a more recent study on virtual sensors [1-10], from how they are implemented; node-level sensor and network-level sensor virtualisation. From this study, open problems and future research directions have also been noted from the study. Additionally, the authors have also been able to classify from the literature (using √ and X to show the presence and absence of a virtual sensor component respectively), whether the identified sensors have intelligent and security components and this is shown in Table 1.

From Table 1, the DIVS [13] is an intelligent sensor that allows multiple logical instances to run simultaneously through node level virtualisation and based on its representation, information security concerns are hardly addressed. Importantly, the study on threats should be more focused on checking the integrity of the information being transmitted from the user to the DIVS service through the UFP that was highlighted in the adversarial threat model. Next, a cloud of virtual sensors by [27] has a sensor implemented at network level virtualisation and this sensor is not intelligent and the security and information privacy concerns are not discussed.

The possible threat area for this sensor is that it lacks security techniques for virtual, intermediate nodes and aggregating data. Also, a virtual sensor as a service [28] that is implemented at network level virtualisation level has not highlighted security and privacy concerns which remains open due to lack of a security component in the cloud which makes it susceptible to attack. Finally, an interactive model based virtual sensors for IoT applications [29] also has open potential threats that need security monitoring of sensor instances coupled with integrity checks. It is worth to mention that what is partial and what is similar on information security and privacy is shown in Table 1 and the discussion has been presented from a cursory investigation.

**Table 1. Virtual sensors implementation level overview and possible threats**

|   | Literature on Virtual Sensors | Dynamic and Intelligent | Node-level Sensor Virtualisation | Network-level Sensor Virtualisation | Information Security & Privacy | Sensor Threat |
|---|---|---|---|---|---|---|
| 1 | Dynamic Intelligent Virtual Sensor (DIVS) [13] | √ | √ | X | X | Integrity of transmitted information, privacy and trust of the transmitting parties |
| 2 | A cloud of virtual sensors [27] | X | X | √ | X | Security of virtual and intermediate nodes and security of aggregating data |
| 3 | Virtual sensor as a service [28] | X | X | √ | X | Lack of security component in the cloud-centric IoT architecture. Lack of sensor activity detection |
| 4 | A location-based interactive model of IoT-sensors [29] | √ | √ | X | X | Secure virtual Sensor instance monitoring and integrity checks |

**VIoT security goals**
The IoT ecosystem which is heterogeneous consists of "things", which also consist of a number of sensors that are able to collect and transmit sensor data based in an IoT environment. The need for access-control infrastructure in IoT has been highlighted as an important approach that can mitigate security breaches and leakage of sensor data. Generally, IoT consists of features that are able to be sensed in a computer network, actuation nodes etc. These "things" can also be monitored within an IoT environment, either in a virtual or physical setting as is highlighted by the IEEE 1451 family of standards and interfaces [30]. The security in IoT context plays a vital role for ensuring the safety of information and the devices within IoT ecosystem. Figure 2 shows the relationship that exists between the sensor data and security goals.

The security layers should be added to the communication and transmission of sensor data and the things that carry data need to have a relationship to the physical devices in order for communication to be complete. Other relevant devices include a data capturing device, sensors and actuators [5]. General IoT communicating devices ensure effective communication over a device that has embedded processing. Consequently, from the perspective of IoT, it is possible to protect sensor data, in memory, at rest, in transit and also end-to-end security from the user to the service to the physical hardware.

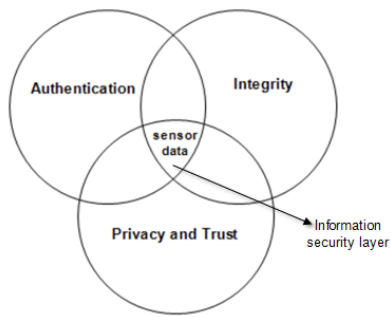

**Figure 2: The relationship between VIoT goals**

Figure 2, shows a representation of the VIoT security goals that need to be achieved by the virtual IoT environment. It is imperative to note that the prime objective of these goals is to ensure the safety of the sensor data and communication. Authenticity is the basic building block for a strong IoT ecosystem while privacy and trust limit the unnecessary exchange of information through proper verifications of the identities of things and users. IoT integrity provides a mechanism of cryptographic protection of sensor data, which provides a strong approach for end-to-end protection of data in IoT environment.

Most of the IoT-based virtual sensors transmit information without necessary safety even though security holds paramount importance. The security and privacy of virtual sensors is a critical issue that at the time of writing this article has not been explored extensively. Disregarding the security of information that is passed by virtual sensors means that the full benefits of IoT cannot be achieved. Additionally, the availability of many IoT communication devices has increased the threat landscape and security risks have increased. Given the increased number of connected devices, the IoT technologies also face formidable security challenges, standardisation issues and communication complexities. Based on the DIVS goals, we have classified threats based on active or passive threats. In this context, active threats are achieved by modifying the functionality of IoT systems while passive threats are through the communication channel. This is shown in Figure 3.

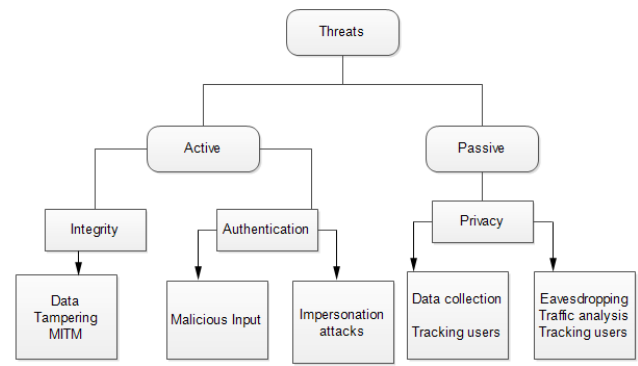

**Figure 3: Mapping DIVS security goals to potential threats**

The DIVS security goals have been mapped to the potential virtual sensor threats. Based on the three goals: integrity, authenticity, privacy and trust, a selected number of threats are mapped to each goal in a generic way. However, each of the IoT supported technology could still face other attacks. For example, Man In The Middle (MITM), data tampering, malicious input and impersonation attacks are categorized as active threats under DIVS integrity and authenticity goals respectively as is shown in Figure 2 previously. Apart from that, data collection, tracking users, eavesdropping, and traffic analysis have been categorised as passive threats under DIVS privacy goal.

That notwithstanding, IoT security technologies that support virtual sensing (See Section 2.3) are mainly constructed to support low-power devices and resource-constrained devices. As a result, the expansion of IoT and the complexity of how information security can be managed keeps changing. It is worth elaborating that there exist other forms of attacks that culminate from IoT communication technologies, which in the long run affect the virtual sensing. For example, privacy of information and are threats that face 5G [14], RFID [15] faces integrity attacks, WSN [16] technology faces Denial of Service (DoS) and Distributed Denial of Service (DDoS) attacks. Low energy Bluetooth [17] faces the threats of blue jacking and blue snarfing,

ZigBee [18] is susceptible to Man in the Middle (MITM) attacks, eavesdropping for Wireless Fidelity (WIFI) [19] and port obscurity for MQTT [20]. Physical and Media Access Control (MAC) attacks for IEEE 802.15.4 [20], DoS and eavesdropping for LoRaWAN [20] and IP spoofing for 6LoWPAN [20]. It is important to note that there may exist many threats as a result and the selected threats have been used for illustrative purposes.

## 5. OPEN SECURITY ISSUES AND RESEARCH DIRECTIONS

In this section, the authors give a discussion on the open security issues in virtual sensors and research directions that are worth taking. The important aspect of the aforementioned concept is the learning/adaptability capability of DIVS to changing environments. VIoT concept which culminates from the susceptibility of virtual sensors to threats in the IoT environment is still an emerging phenomenon given that still this research area is less explored at the time of writing this paper. Furthermore, the rise of the threats in IoT has been as a result of increased sensor technologies, increased number of devices, and increased amount of information that are produced by these devices. For example, the DIVS concept puts forward a virtual sensor that is deployed in the IoT environment to accumulate sensor data in an environment that has a multitude of sensor threats. The authors explore the following issues and research directions:

- ***Protecting the communication channel from virtual sensor:*** Little focus has been put on how one can ensure that the operations of virtual sensors are able to overcome integrity threats. It is important for IoT tool designers to be able to design tools that are able to identify threats that relate to malicious configurations that can hamper or compromise the integrity of sensor data. Further research should also focus on creating dynamic intelligent virtual sensor configurations that are tamper free from alterations and modification of communication process. Through this, the accuracy of the online learning process can be guaranteed.

- ***Virtual sensor resilience:*** Generally virtual sensors form part of the IoT system at large and it is important to ensure that if the virtual nodes are compromised, the IoT's functionality should continue to operate. It is vital that the compromised nodes are identified, isolated and reported. Further research should be focused on not being able to change the existing functionality of the IoT system in case vulnerabilities or an attack is detected, but also for the virtual sensor to continue operating with a high level of accuracy.

- ***Privacy:*** There is a need for ensuring that privacy enhancing technologies are employed to IoT generated data that moves across connected things through data segregation and separation. Research directions should be more focused on protection through aggregation of data through the use of different multiple levels of secure access in order to prevent unauthorised access to individual data even when security cameras are used.

- ***Virtual sensor attribution:*** While it remains important to discover what IoT device may be attributed to a particular threat or attack, it is also important to focus on virtual sensor attribution. This is mainly because when virtual sensors may be used as attack objects, they have a possibility of interfering with an IoT environment either actively or passively [33].

## 6. DISCUSSIONS

We revisit the hypothetical scenario that has been highlighted in Section 1 of this paper. The scenario mainly focused on the drawbacks that are achieved as a result of attacks on integrity, confidentiality and authenticity as a result of the existing threats on virtual sensors. Given how virtual sensing is achieved in an IoT environment, succeeding with these attacks is considered a serious breach of security techniques that can easily compromise a whole IoT environment. *X, the malicious user* from the hypothetical scenario (section 1) has been able to achieve malicious goals through spoofing and the threats are realised as soon as *X* is able to shut down the smart cameras and mount illegal sensor nodes through a rented VM. Consequently, given that the scenario pinpoints the failure on how security could be enforced, and a success on the malicious goals by *X*, we review the security goals that the DIVS which has been used as a basis of study in this paper is meant to achieve. It is therefore, an important measure to ensure the adoption of an IoT architecture with security capabilities for the DIVS/virtual sensors.

The security techniques that can help to protect virtual sensors in the IoT environment should mainly focus on how the information that is passed between sensors and the environment is being sensed. While the internet and communication carry much importance, it is also important to say that it acts as a safe haven for attackers and it could be used to propagate attacks. If we revisit the DIVS concept (section 2.2), it is an example of a virtual and dynamic intelligent sensor that needs information security protection techniques that can safeguard information that is being relayed. VIoT introspection attempts to do an extensive exploration on how susceptible the virtual sensors are to threat tribulations in an IoT environment and also it shows the drawbacks that this may have to IoT communication technologies that has been shown in Table 3 of this article. A more realistic approach that highlights the mechanism of hardening the virtual sensing in IoT is the use of four-layer IoT architecture that has the support of security recommendations that are aimed at protecting IoT communications [31], [32]. This security recommendation span across four layers namely the application layer (1) that ensures there is proper authentication/key agreement and privacy during message passing. Then this is followed by a support layer (2) that ensures that there is secure cloud computing in case resources are being shared, then network layer (3) that supports identity authentication and encryption approaches. Lastly, the perception layer (4) that supports encryption and key agreement in order to support the sensor data. Given that there is direct information passing at the DIVS, it makes the threats to be likely to be forthcoming and based on these, the authors echo the importance, that the four-layer IoT security architecture, may play as far as this information security of DIVS is concerned. If we revisit the user-feedback process (UFP)/flow in the DIVS, information is expected to be transmitted or sent by authentic users from the application layer where their authenticity and privacy can be enhanced. It is important to say that the functionality of virtual sensors allows logical instances to be used on an on-demand basis and this raises the question of the logical instances being threats to other virtual sensors/instances. In this case, the virtual instance could be used to propagate attacks, where it could be possible for the attacker to use an instance and then shut down the virtual component or change the location of the virtual component.

## 7. CONCLUSION AND FUTURE WORK

This paper has introduced the concept of VIoT introspection and the authors have concentrated on giving discussions on the need for introducing information security layers in virtual sensing. This study gives a comprehensive overview at virtual sensors from an information security perspective. It is the authors' belief that the study will have a broad impact as far as virtual sensor threats are concerned. While this is work in progress, future work is aimed at creating a real-time attack detection VIoT test-bed to be able to identify and mitigate the virtual IoT sensor threats.

## ACKNOWLEDGEMENT

This research was partially funded by The Swedish Knowledge Foundation through the Internet of Things and People grant number 20140035.